\documentclass[11pt, oneside]{article}

\usepackage[english]{babel}
\usepackage{geometry}
\usepackage{graphicx}
\usepackage{tabularx}
\usepackage{amssymb}
\usepackage{amsmath}
\usepackage{booktabs}
\usepackage{xltabular}
\usepackage[hyphens]{url}
\usepackage{hyperref}
\hypersetup{breaklinks=true,hidelinks}
\usepackage[backend=bibtex,style=nature]{biblatex}
\usepackage{csquotes}
\usepackage{cleveref}
\usepackage{siunitx}
\usepackage{geometry}
\usepackage{multirow}
\usepackage[dvipsnames]{xcolor}
\definecolor{tumblue}{rgb}{0,0.396078431372549,0.741176470588235}

\usepackage{xspace}
\usepackage{caption}
\usepackage{subcaption}
\usepackage{authblk}
\usepackage{mhchem}

\usepackage{marvosym}

\usepackage{sectsty}

\sectionfont{\color{tumblue}\selectfont\sffamily}
\subsectionfont{\color{tumblue}\selectfont\sffamily}
\subsubsectionfont{\color{tumblue}\selectfont\sffamily}
\paragraphfont{\selectfont\sffamily}
\subparagraphfont{\color{gray}\selectfont\sffamily}

\definecolor{halfgray}{gray}{0.55}

\IfFileExists{./fonts/Heliotrope-Bold.otf}{
    \usepackage{fontspec}
    \setmainfont{Heliotrope}[Extension = .otf, UprightFont = *-Regular, ItalicFont = *-Italic, BoldFont=*-Bold, Path = ./fonts/]
}{
    \usepackage{biolinum}
    
}

\makeatletter
\renewbibmacro*{textcite}{%
  \iffieldequals{namehash}{\cbx@lasthash}
    {\usebibmacro{cite:comp}}
    {\usebibmacro{cite:dump}%
     \ifbool{cbx:parens}
       {\bibclosebracket\global\boolfalse{cbx:parens}}
       {}%
     \iffirstcitekey
       {}
       {\textcitedelim}%
     \usebibmacro{cite:init}%
     \ifnameundef{labelname}
       {\printfield[citetitle]{labeltitle}}
       {\printnames{labelname}}%
     \addspace
     \ifnumequal{\value{citecount}}{1}
       {\usebibmacro{prenote}}
       {}%
     \mkbibsuperscript{\usebibmacro{cite:comp}}%
     \stepcounter{textcitecount}%
     \savefield{namehash}{\cbx@lasthash}}}
\makeatother

\usepackage{lipsum}
\usepackage[labelfont=bf]{caption}

 \usepackage{microtype}

\usepackage{tikz}
 \usetikzlibrary{positioning,fit,backgrounds,arrows.meta,calc}
\usepackage{xcolor}

\definecolor{dangercolor}{RGB}{180,30,30} 
\definecolor{chemcolor}{gray}{0.2}        
\definecolor{metacolor}{gray}{0.2}        

\usepackage{credits}
\usepackage{orcidlink}
\usepackage{showyourwork}

\title{\textsf{Clever Materials: When Models Identify Good Materials for the Wrong Reasons}}

\author[1,2, 3, 4, \Letter]{Kevin~Maik~Jablonka~\orcidlink{0000-0003-4894-4660}}

\affil[1]{Laboratory of Organic and Macromolecular Chemistry (IOMC), Friedrich Schiller University Jena, Humboldtstrasse 10, 07743 Jena, Germany}
\affil[2]{Helmholtz Institute for Polymers in Energy Applications Jena (HIPOLE Jena), Lessingstrasse 12-14, 07743 Jena, Germany}
\affil[3]{Center for Energy and Environmental Chemistry Jena (CEEC Jena), Friedrich Schiller University Jena, Philosophenweg 7a, 07743 Jena, Germany}
\affil[4]{Jena Center for Soft Matter (JCSM), Friedrich Schiller University Jena, Philosophenweg 7, 07743 Jena, Germany}

\affil[\Letter]{\texttt{mail@kjablonka.com}}

\let\oldabstract\abstract
\let\oldendabstract\endabstract
\makeatletter
\renewenvironment{abstract}
{%
               {\list{}{\addtolength{\leftmargin}{-3em} 
                        \listparindent 1.5em%
                        \itemindent    \listparindent%
                        \rightmargin   \leftmargin%
                        \parsep        \z@ \@plus\p@}%
                \item\relax}%
               {\endlist}%
\oldabstract}
{\oldendabstract}
\makeatother

\begin{document}

\maketitle
\vspace*{-3em}
\begin{figure}[h]
	\centering
	\includegraphics[width=.5\textwidth]{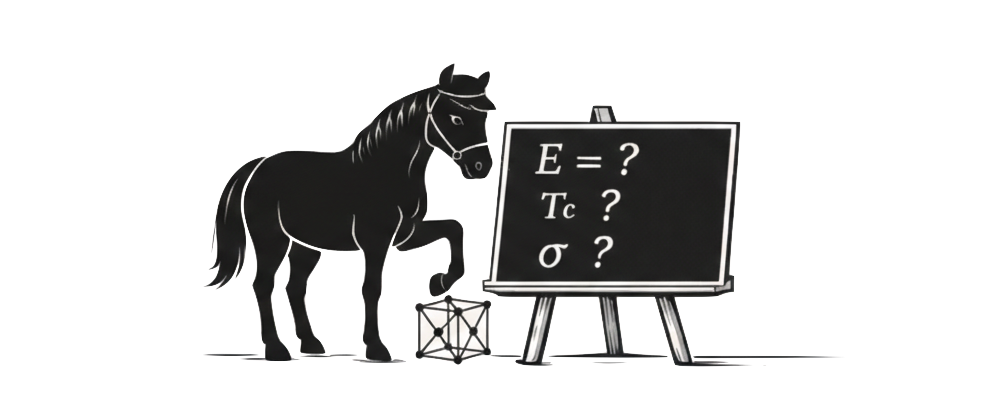}
\end{figure}

\begin{abstract}
Machine learning can accelerate materials discovery. Models perform impressively on many benchmarks. 
However, strong benchmark performance does not imply that a model learned chemistry. I test a concrete alternative hypothesis: that property prediction can be driven by bibliographic confounding.
Across five tasks spanning MOFs (thermal and solvent stability), perovskite solar cells (efficiency), batteries (capacity), and TADF emitters (emission wavelength), models trained on standard chemical descriptors predict author, journal, and publication year well above chance. When these \emph{predicted} metadata (\enquote{bibliographic fingerprints}) are used as the sole input to a second model, performance is sometimes competitive with conventional descriptor-based predictors.
These results show that many datasets do not rule out non-chemical explanations of success. Progress requires routine falsification tests (e.g., group/time splits and metadata ablations), datasets designed to resist spurious correlations, and explicit separation of two goals: predictive utility versus evidence of chemical understanding.
\end{abstract}

\section{Introduction}

Machine learning holds both promise and peril for materials discovery. 
The promise lies in its capacity to uncover structure-property relationships too subtle or complex for human recognition.\autocite{hardtrecht2022patterns} 
The peril emerges when models excel by exploiting spurious patterns that collapse under new conditions.\autocite{Lones2024}

This phenomenon---impressive demonstration performance masking fundamental brittleness---represents a classic failure mode in pattern recognition, epitomized by the horse \enquote{Clever Hans}.\autocite{Lapuschkin2019} 
Clever Hans appeared to perform arithmetic calculations, fooling audiences until careful investigation revealed he was simply reading cues from his questioners. 

Modern machine learning exhibits analogous vulnerabilities. 
Computer vision models exploit spurious correlations---skin color in medical diagnosis,\autocite{pooch2019trust} background textures in animal classification\autocites{xiao2020noise}---achieving high accuracy through irrelevant shortcuts.\autocite{Brown2023, Howard2021} 
A recent report from Leash Biosciences suggest similar risks in chemical property prediction: models achieve surprising accuracy at predicting compound performance, potentially using authorship as a proxy for bioactivity rather than learning meaningful chemistry.\autocite{leash} 

\begin{figure}[!htb]
	\centering 
	\input{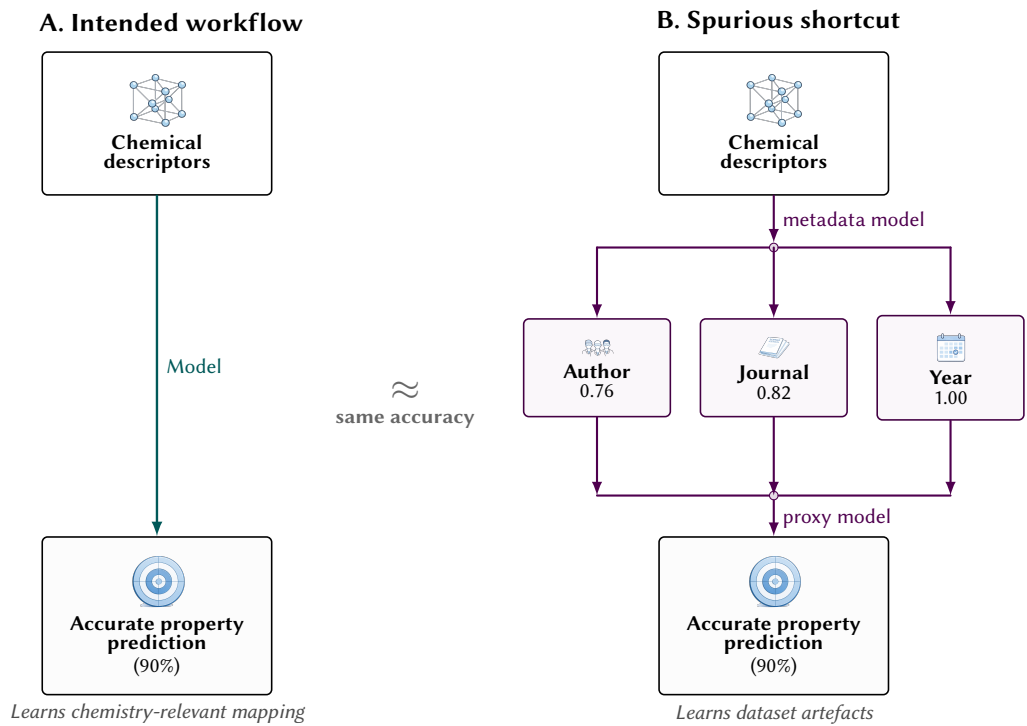}
	\caption{\textbf{In machine learning, we often do not test competing hypotheses of how a model might obtain its answers.} Machine learning models in materials science are trained to map material (descriptors) to property predictions. Models have much flexibility in how they learn this map from data. Ideally, they discover robust and meaningful structure-property relationships that also generalize in new settings. This, however, is not guaranteed. Models might also exploit other patterns in the data as a shortcut to a prediction (purple arrows). For instance, it might be easy for the model to spot what researchers produced a given material or in what journal it has been published. Based on those inferences, it might deduce property \enquote{guesses} (as the knowledge of the research group or the publication time can be correlated to the property). The model thus might learn to make good predictions for the wrong reasons. This is known as the \enquote{Clever Hans} effect. The scientific method asks us to test if such alternative patterns can explain good model performance. This is seldom done. In this work, I do it for a few case studies. }	
	\label{fig:clever_hans}
\end{figure}

Materials science offers abundant opportunities for such proxy learning (\Cref{fig:clever_hans}). 
Research groups develop specialized expertise---some focus on solar cell stability optimization, others on MOF synthesis strategies. Laboratory names often appear in framework designations (UiO-66, MIL-101), creating direct author-material associations. 
Fields evolve through paradigm shifts\autocite{kuhn1997structure} and with publication bias this might embed temporal signatures that models might exploit rather than learning fundamental chemistry.

This work systematically investigates whether commonly used materials datasets are vulnerable to such proxy learning. 
I test a simple hypothesis: can models predict material properties using only (predicted) bibliographic metadata---author names, publication years, journal venues---rather than chemical descriptors? 
The answer is, in some cases, yes. 
Models predict bibliographic metadata from chemical descriptors with surprising accuracy, and models using only these predicted \enquote{bibliographic fingerprints} can match the performance of conventional structure-property approaches. 

These findings reveal how easily we can be misled about what our models actually learn. 
More critically, they expose a systematic gap in how we validate machine learning in materials science: we optimize performance without testing competing hypotheses about \emph{why} models work.

\section{Results}

I audit bibliographic shortcut learning in five materials prediction tasks: MOF thermal stability, MOF solvent stability, perovskite solar-cell efficiency, battery capacity, and TADF emission wavelength. 
For each task I compare three model classes under identical cross-validation splits: (i) a \emph{direct} model mapping chemical descriptors to properties, (ii) a \emph{metadata} model mapping descriptors to bibliographic variables (authors, journals, years), and (iii) a \emph{proxy} model mapping \emph{predicted} bibliographic variables to properties.
If the proxy model approaches the direct model performance, good performance due leveraging meaningful chemical patterns is indistinguishable from good performance due to leveraging shortcuts.

\subsection{MOF Thermal Stability}

Thermal stability represents a critical constraint for MOF applications across gas storage, catalysis, and separation processes.\autocite{Kalmutzki2018} 
\textcite{Nandy2022} systematically extracted decomposition temperatures from thermogravimetric analyses reported in the literature, creating opportunities to model thermal stability as either a continuous property (regression) or a discrete stability class (classification).\autocite{Nandy2021}

\begin{figure}[htb]
	\centering 
	\includegraphics[width=\textwidth]{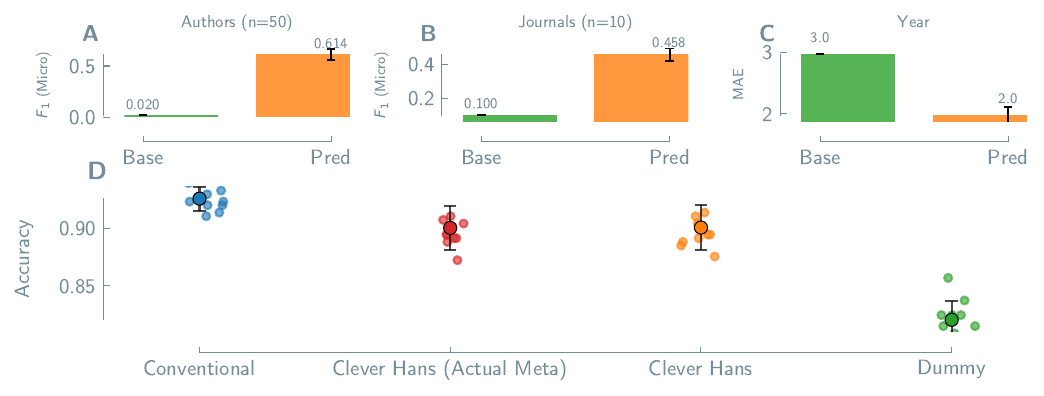}
	\caption{\textbf{For the classification task of membership in the top-10\% of thermally stable MOFs, one can be fooled (by Clever Hans effects).}  \textbf{a} The model predicts the authors of the associated paper with high accuracy, much better than a random baseline. \textbf{b} This also holds for predicting in which journal the entry was published or the year in which the paper was published \textbf{c}. Using the predicted bibliographic information, a model can predict with high accuracy if the MOF belongs to the top-10\% thermally stable ones. However, the effect is smaller --- or not even there --- if analyzed under a different metric or for a regression setting (see \Cref{sec:appx-mof-thermal-stability}). The dummy baseline for classification is a stratified random sampling (using the empirical probabilities from the training dataset) and the mean prediction for the regression case.}
	\label{fig:mof_thermal_stability}	
	\script{analyze-mof-thermal-top10.py}	
\end{figure}

\Cref{fig:mof_thermal_stability} demonstrates measurable proxy learning: structural descriptors predict bibliographic information significantly above baseline, enabling top-10\% thermal stability classification with an accuracy of %
  0.901 \unskip\label{output/mof_thermal_top10_indirect_accuracy.txt}\unskip%
, approaching that of conventional structure-property models (%
  0.923 \unskip\label{output/mof_thermal_top10_direct_accuracy.txt}\unskip%
).
However, \Cref{fig:mof_thermal_stability_metric_impact} shows that the measured effect size depends on the metric one uses to compare models.

\subsection{MOF Solvent Stability}

Solvent removal stability poses an equally critical challenge for MOF deployment. 
Many frameworks collapse when synthesis solvents are removed during activation, limiting their practical utility. 
Using a text-mined dataset and descriptors \textcite{Nandy2022}, I tested whether proxy signals could predict solvent stability outcomes.

\begin{figure}[htb]
	\centering
	\includegraphics[width=\textwidth]{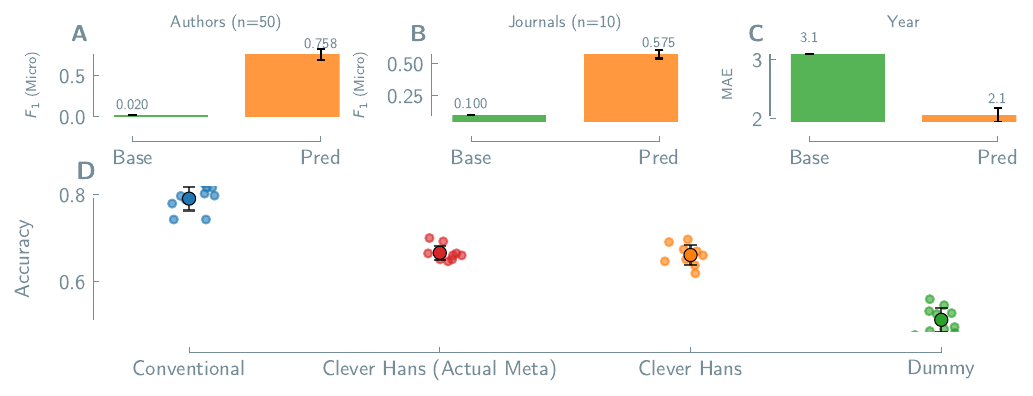} 
	\caption{\textbf{Measuring performance using accuracy, \enquote{Clever Hans models} can achieve a surprisingly good performance in predicting solvent removal stability of MOFs.} \textbf{a} MOF descriptors can be used to predict author information meaningfully better than a random baseline. \textbf{b} The journal in which a given MOF stability result has been published can also be predicted with non-trivial accuracy. \textbf{c} The publication year can be predicted with a mean average error of around two years. \textbf{d} A model trained on predicted bibliographic information (actual bibliometric and predicted one) can achieve non-trivial accuracy in correctly predicting the solvent stability of MOFs. \Cref{fig:mof_solvent_removal_stability_metric_impact} shows the performance measured with other metrics.}
	\label{fig:mof_solvent_stability}
	\script{analyze-mof-solvent-stability.py}
\end{figure}

\Cref{fig:mof_solvent_stability} shows that MOF structural descriptors predict bibliographic metadata with moderate accuracy. The proxy model achieves classification accuracy of %
  0.655 \unskip\label{output/mof_solvent_indirect_accuracy.txt}\unskip%
, above baseline but below direct structure-property approaches, indicating partial Clever Hans susceptibility.

\subsection{Perovskite Solar Cell Efficiency}

Perovskite solar cells are another way in which materials scientists aim to have a positive impact on the energy transition.\autocite{CorreaBaena2017}
A metric of central importance here is the power conversion efficiency (PCE). 
It was mined by \textcite{Jacobsson2021} in a manual approach and by \textcite{shabih2026autonomous} in an automated one with large language model-based data extraction.\autocite{SchillingWilhelmi2025}

\begin{figure}[htb]
	\centering 
	\includegraphics[width=\textwidth]{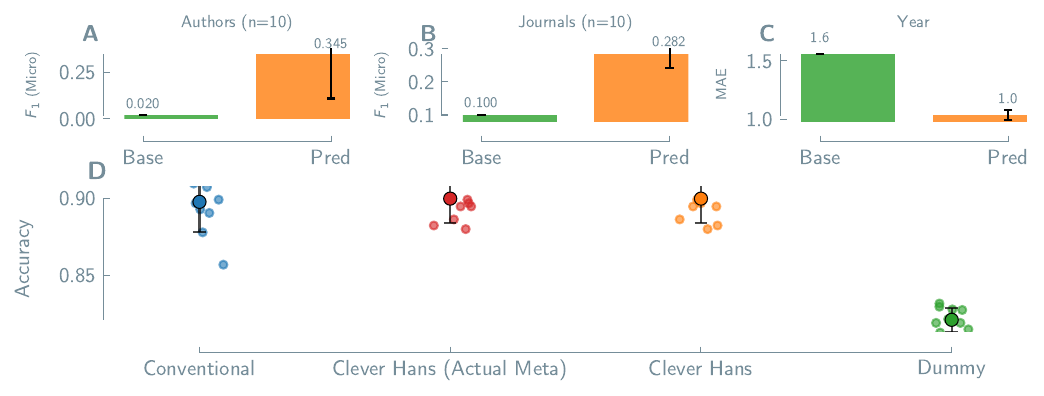}
	\caption{\textbf{\textbf{Clever Hans} models can achieve a high accuracy in predicting if a perovskite material is within the top-10\% most efficient absorbers.} \textbf{a} Composition description can predict which of the ten most prolific authors has been reporting a composition with a micro $F_1$ score of \protect%
  \input{output/perovskite_top10_author_f1_micro.txt}\unskip\label{output/perovskite_top10_author_f1_micro.txt}\unskip%
. \textbf{b} The journal in which a device has been published can be predicted with a similar performance. \textbf{c} The publication year can be predicted meaningfully better than with a mean baseline. \textbf{d} A model trained on this predicted bibliometric information can achieve an accuracy in classifying if an aborber belongs to the top-10\% most efficient ones with an accuracy indistinguishable from the model directly trained on composition features.}
	\label{fig:perovskite_pce_classification}
	\script{analyze-perovskites-top10.py}
\end{figure}

\Cref{fig:perovskite_pce_classification} demonstrates that perovskite composition descriptors predict bibliographic information with meaningful accuracy. 
The model achieves a micro-averaged $F_1$-score of %
  0.318 \unskip\label{output/perovskite_top10_author_f1_micro.txt}\unskip%
 for the 10 most prolific authors, indicating detectable authorship signatures in the chemical data. 
Similar performance occurs for journal and publication year prediction.

The proxy model achieves classification accuracy of %
  0.900 \unskip\label{output/perovskite_top10_indirect_accuracy.txt}\unskip%
, comparable to direct composition-property models (%
  0.899 \unskip\label{output/perovskite_top10_direct_accuracy.txt}\unskip%
) for identifying top-10\% efficient devices. This suggests potential reliance on author expertise patterns or temporal trends rather than composition-efficiency relationships. 
While also this effects depends on the metric one uses, the results highlight that the use of a simple naïve dummy baseline (e.g., predicting prior probabilities) is not enough to demonstrate chemically meaningful learning.

\subsection{TADF Emitter Properties}

Thermally activated delayed fluorescence (TADF) is one mechanism to improve the efficiency of organic light-emitting diodes (OLED)s.\autocite{Liu2018} 
The maximum emission wavelength is one important performance metric that is optimized for these materials. \textcite{Huang2024} text-mine this property using ChemDataExtractor.\autocite{Swain2016, Mavrai2021}

Molecular descriptors and fingerprints serve as features for both conventional and proxy models predicting maximum emission wavelength. 

\begin{figure}
	\centering
	\includegraphics[width=\textwidth]{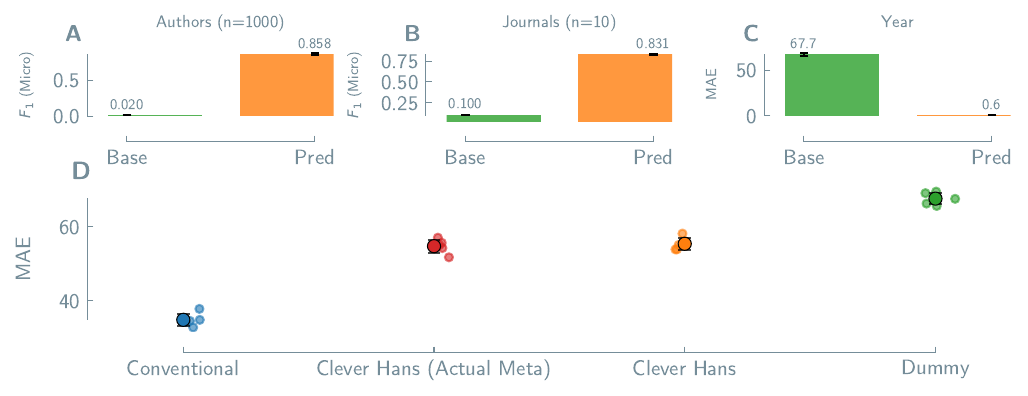}
	\caption{\textbf{\enquote{Clever Hans models} perform worse than \enquote{conventional} models but better than simple baselines in predicting the maximum emission wavelengths of TADF molecules.} \textbf{a} Using molecular descriptors, the model can predict the the authors of the paper describing a material with an accuracy of \protect%
  \input{output/tadf_best_meta_accuracy.txt}\unskip\label{output/tadf_best_meta_accuracy.txt}\unskip%
\%  among the 1000 most profilic authors. \textbf{b} The model achieves an even higher performance in predicting in which of the ten most common journals a given entry has been published. \textbf{c} The publication year, too, can be predicted with a high performance just based on composition descriptors. \textbf{d} Using predicted bibliometric information, one can achieve a mean absolute error between the \enquote{conventional} and naïve baseline models. The performance using actual bibliometric information is not much different from using the correct one.}
	\label{fig:tadf_main}
	\script{analyze-tadf.py}
\end{figure}

\Cref{fig:tadf_main} demonstrates that TADF molecular descriptors enable moderate bibliographic prediction. 
The proxy model achieves a mean average error for maximum emission wavelength prediction between conventional and baseline approaches, indicating limited but detectable Clever Hans effects.

\subsection{Battery Capacity}

For sustainability, energy does not only need to be converted, but also stored. For this, batteries are important. \textcite{Huang2020} text-mined battery materials alongside performance metrics. 

\begin{figure}[htb]
	\centering 
	\includegraphics[width=\textwidth]{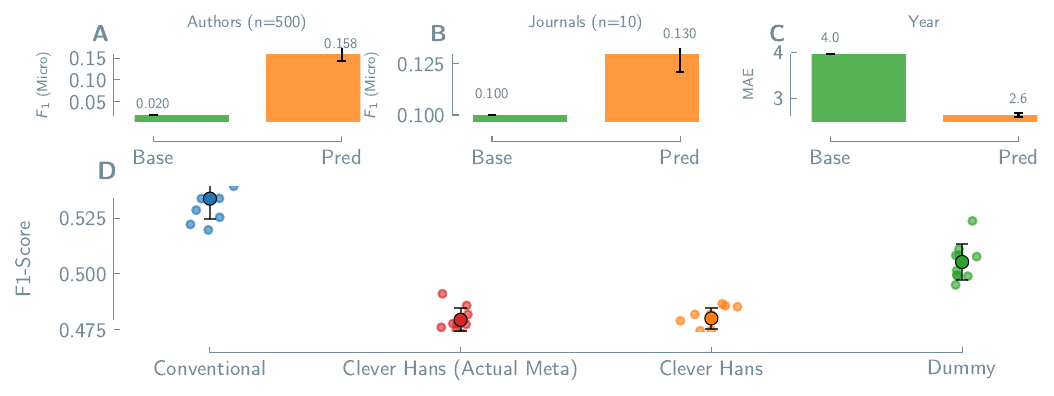}
	\caption{\textbf{\enquote{Clever Hans models} do not achieve a good performance in predicting battery capacity.} \textbf{a} Composition descriptors can be used to determine which of the 500 most prolific authors reported this material with a mediocre $F_1$ score. \textbf{b} The journal in which a material has been reported can be predicted with a similar performance. \textbf{c} The publication year can be predicted better than with the mean baseline. \textbf{d} \enquote{Clever Hans} models are not distinguishable from mean predictions (\enquote{Dummy}) for the prediction of the capacity of battery materials.}
	\label{fig:battery_capacity}
	\script{analyze-batteries-top10.py}
\end{figure}

\Cref{fig:battery_capacity} shows that battery composition descriptors exhibit limited proxy learning capability. Author prediction achieves a mediocre $F_1$-score, while publication year prediction performs moderately above baseline. The resulting proxy model does not perform better than simply always predicting the mean. 

This null result is scientifically valuable as it demonstrates that Clever Hans effects are not inevitable artifacts of machine learning, but rather depend on dataset construction and domain characteristics.

\subsection{Overall Effects}

Clever Hans effects vary significantly across material domains and prediction tasks. Proxy models achieved competitive performance in perovskite efficiency classification and MOF thermal stability, moderate effects in TADF wavelength prediction and MOF solvent stability, and negligible effects in battery capacity prediction. Effect detectability depends critically on evaluation metrics and baseline selection. In most cases, the proxy model based on the predicted literature data was a stronger baseline than the a simple baseline on prior probabilities or mean predictions.

\section{Discussion}

Machine learning has transformed materials discovery,\autocite{Moosavi2020, Saal2020, Gubernatis2018} but the findings here highlight an important caveat: we often fail to rigorously test alternative hypotheses for why our models perform well. 
The scientific method demands that we actively seek to falsify our hypotheses,\autocite{Platt1964, popper2005logic, Chamberlin1965} yet in machine learning, we tend to focus on optimizing performance rather than exploring competing explanations.

The Clever Hans effect represents just one class of alternative hypotheses we should systematically explore. 
When we claim that models \enquote{learn meaningful chemistry,} we must test whether simpler explanations --- such as shortcuts via author identity or publication date --- could account for the observed performance.\autocite{Chuang2018} 
This requires a shift from asking \enquote{does this model work?} to \enquote{why does this model work, and what are all the ways it could be wrong?}

The space of potential confounders is vast and often non-obvious. Beyond the shortcuts via meta-information investigated in this work, models might exploit dataset construction artifacts, measurement biases, or many other spurious effects.\autocite{Zhou2025, Jones2024} 
Systematically exploring these alternatives is computationally intensive but crucial for scientific rigor.

LLM-based agents might offer a promising approach to automate this exploration.\autocite{Ramos2025, alampara2025general, narayanan2024aviary0, ghareeb2025robin0, mitchener2025kosmos0, ghafarollahi2025sciagents} 
These systems could generate and test competing hypotheses in parallel, exploring the space of potential explanations more thoroughly than human researchers typically manage. 
Such agents could serve as \enquote{devil's advocates,} systematically challenging our assumptions about why models succeed.

\subsection{Toward Robust Materials Data Infrastructure}

Another angle is to reconsider how we generate, curate, and share materials data. 
The field needs coordinated infrastructure and institutions that prioritize diversity and robustness over convenience. \autocite{Krishnan2025}
Convenience and short-term reward are often too easy and compelling to optimize for due to collective action problems trapping an ecosystem in a suboptimal state, where every actor knows that changes would be needed, but no one wants to make the first move.\autocite{Nielsen2020-rr} 

Most automated screening approaches optimize specific objectives using limited building blocks, which creates exactly the kind of proxy signals that models learn to exploit.  Human researchers, too,  are biased in how they explore chemical space.\autocite{Jia2019} In addition,  publication bias favors positive results, creating temporal performance trends.\autocite{Jablonka2022}
Organizations that can generate diverse data at scale---potentially focused on the lowest cost per reproducible data point rather than pushing particular research agendas---might help to address this problem. 
Other options could be explicit quantification of author/group/temporal distributions, correlation analyses between metadata and properties as \enquote{dataset nutrition labels}, or adversarial dataset construction by deliberately designing datasets to resist spurious correlations.

To enable others to do participate in testing alternative hypothesis for model performance, access to data and code is obviously a prerequisite. But one could also envision that some of these tests might require new experiments --- which could be facilitated using infrastructure as a service or incentivized using \enquote{bug bounties} for research papers, models, or datasets. We need to accept that receiving feedback --- even if it is pointing out a mistake in our own work --- is a gift.

\section{Conclusions}

Model evaluation has always been challenging in materials science.\autocite{Alampara2025} 
We have developed increasingly sophisticated techniques to address this: time-based splitting,\autocite{sheridan2013time, Landrum2023} scaffold splits, leave-one-cluster-out cross-validation,\autocite{Durdy2022, Meredig2018} cluster-based splits,\autocite{guo2024scaffold} and property-based splits.\autocite{Jablonka2023, kunchapu2025polymetrix} 
In some domains, even challenges have been organized.\autocite{Moult2005, Tetko2024, Llinas2019}
Each technique revealed new ways that models could fail to generalize, forcing us to be more rigorous in our evaluation practices.

This work highlights yet another layer of complexity. 
Across five materials prediction tasks, I find that bibliographic signals can be predicted from chemical descriptors and can sometimes support property prediction at near-conventional performance. The effect is heterogeneous. It is strong in some top-10\% classification settings, moderate in others, and absent for battery capacity in this study. This heterogeneity implies that shortcut learning is not inevitable, but also not reliably detected by standard validation alone.
The practical implication is that benchmark accuracy should be treated as ambiguous evidence unless competing shortcut hypotheses are actively tested. Routine audits—metadata ablations, group/time splits, and explicit effect-size reporting—would make it harder for spurious correlations to masquerade as chemical insight, while preserving the legitimate value of models that are \emph{only} intended as in-distribution predictors.


\section{Methods}

\subsection{Clever Hans Analysis Framework}

I implemented a systematic framework to quantify Clever Hans effects in materials property prediction. For each dataset, I trained three types of models: (1) conventional models that predict material properties directly from chemical descriptors, (2) indirect models that first predict meta-information (author identity, journal, publication year) from the same descriptors and then use these predictions to estimate material properties, and (3) dummy baselines using stratified sampling for classification or mean prediction for regression.

The indirect prediction approach tests whether meta-information contains sufficient signal to achieve competitive prediction performance. If models can predict material properties as accurately using only proxy information as using chemical descriptors, this indicates potential Clever Hans effects in the dataset.

\subsection{Model Architecture and Training}

All models used gradient boosting, implemented with \texttt{LightGBM}\autocite{lightgbm} with default hyperparameters. 

\subsection{Cross-Validation Protocol}
I used 10-fold cross-validation with random shuffling unless otherwise mentioned. Individual dots in swarm plots indicate the performance on the individual folds. 
For each of the 10 cross-validation folds, I maintain strict separation between training and testing phases: In the training phase, three models are trained simultaneously on the same training data: (1) the conventional model learns to map chemical descriptors directly to target properties, (2) the meta-prediction model learns to predict bibliographic information (authors, journals, publication years) from chemical descriptors, and (3) the proxy model learns to map predicted bibliographic information to target properties.
In the testing phase, the conventional model predicts properties from descriptors on the held-out fold. 
The meta-prediction model generates bibliographic predictions for the test materials, and the proxy model uses the predicted bibliographic data (not ground-truth metadata) to predict properties.
This protocol ensures that proxy models cannot access ground-truth bibliographic information during testing, only their own uncertain predictions. The comparison thus reflects realistic deployment scenarios where future materials would lack known authorship or publication context.

\subsection{Datasets and Feature Engineering}

\subsubsection{Battery Dataset}
I obtained the battery dataset from \textcite{Huang2020}
The battery dataset contained %
  33701 \unskip\label{output/battery_top10_dataset_size.txt}\unskip%
 entries with %
  273 \unskip\label{output/battery_top10_n_features.txt}\unskip%
 chemical descriptors.

\subsubsection{Perovskite Dataset} 
I obtained the perovskite dataset from \textcite{shabih2026autonomous}, which is based on \textcite{Jacobsson2021}.
The perovskite dataset contained %
  4753 \unskip\label{output/perovskite_dataset_size.txt}\unskip%
 entries with %
  273 \unskip\label{output/perovskite_n_features.txt}\unskip%
 descriptors.

\subsubsection{MOF Datasets}
I obtained the MOF datasets from \textcite{Nandy2022}. The dataset already contains precomputed features such as revised autocorrelation functions.\autocite{Moosavi2020} 
The MOF thermal stability dataset contained %
  3131 \unskip\label{output/mof_thermal_dataset_size.txt}\unskip%
 entries with %
  174 \unskip\label{output/mof_thermal_n_features.txt}\unskip%
 structural and chemical descriptors. 
The MOF solvent stability dataset contained %
  2179 \unskip\label{output/mof_solvent_dataset_size.txt}\unskip%
 entries with %
  174 \unskip\label{output/mof_solvent_n_features.txt}\unskip%
 descriptors.
I used the MOF descriptors provided by \textcite{Nandy2022}. 

\subsubsection{TADF Dataset}
I obtained the TADF dataset from \textcite{Huang2024}. 
The TADF dataset contained %
  2119 \unskip\label{output/tadf_best_dataset_size.txt}\unskip%
 entries with %
  2265 \unskip\label{output/tadf_n_features.txt}\unskip%
 molecular descriptors. As inputs for the models, I used compositional descriptors.

\subsection{Chemical Descriptor Generation}

For datasets containing molecular or compositional information, I generated chemical descriptors to serve as baseline features for property prediction. 

\subsubsection{Molecular Descriptors from SMILES}
For datasets (TADF) with SMILES (Simplified Molecular-Input Line-Entry System) strings,\autocite{Weininger1988} I computed molecular descriptors using RDKit \autocite{rdkit}. The molecular feature set included:

\begin{itemize}
    \item \textbf{2D descriptors}: All available RDKit molecular descriptors ($\sim$200 features), including molecular weight, LogP, topological polar surface area, number of aromatic rings, hydrogen bond donors/acceptors, and rotatable bonds.
    \item \textbf{Fingerprints}: 2048-bit circular fingerprints with radius 2, capturing local chemical environments and structural motifs.
\end{itemize}

Molecules were parsed from SMILES strings, and invalid or unparseable structures were excluded. 
\subsubsection{Composition Descriptors}
For datasets with chemical formulas (battery materials, perovskites), I computed composition-based descriptors using matminer \autocite{matminer}. The composition feature set included:

\begin{itemize}
    \item \textbf{Element properties}: Elemental statistics (mean, standard deviation, range) for atomic properties, including atomic radius, electronegativity, ionization energy, and electron affinity using the Magpie preset \autocite{ward2016general}.
    \item \textbf{Stoichiometric features}: Composition statistics including element fractions, number of components, and chemical complexity metrics.
    \item \textbf{Meredig descriptors}: Extended element property statistics including orbital contributions and chemical bonding characteristics \autocite{meredig2014combinatorial}.
\end{itemize}

Chemical formulas were parsed using pymatgen \autocite{pymatgen}, and compositions that could not be parsed were excluded from analysis. 

\subsubsection{Feature Processing}
Generated descriptors were processed to handle missing values and ensure numerical stability for gradient boosting models. Features with excessive missing values ($>$50\%) were excluded, and remaining missing values were imputed with feature medians. Additional preprocessing included clipping extreme values to prevent numerical overflow and replacing infinite values with conservative bounds.

\subsection{Meta-Information Extraction}

I enriched the datasets with publication meta-information using the Crossref API to retrieve bibliographic data, including author names, journal titles, and publication years. I created binary features indicating the presence of the top-$N$ most frequent authors and journals in each dataset, where $N$ was varied across 10, 50, 100, and 500 (or maximum available).

\subsection*{Data and Code Availability}

To ensure reproducibility, this manuscript was generated using the \href{https://show-your.work/en/latest/}{\showyourwork} framework.\autocite{Luger2021}
The code to rebuild the paper (including code for all figures and numbers next to which there is a GitHub icon) can be found at \url{\GitHubURL}.

\section*{Acknowledgement}
This work was supported by the Carl Zeiss Stiftung. The author is a member of the NFDI consortium FAIRmat - Deutsche Forschungsgemeinschaft (DFG) - Project 460197019.

\section*{Declaration of Generative AI and AI-assisted Technologies in the Research and Writing Process}
I used Anthropic's Claude models as \enquote{copilot} in code development. I also used those models to improve language and readability. After using this service, I reviewed and edited the content as needed and take full responsibility for the content of the publication.

\printbibliography

\pagebreak

\appendix

\section{Detailed Results}
In this section, I show performance for metadata and property prediction in more detail.

\subsection{MOF Solvent Removal Stability}
\begin{figure}[htb]
	\includegraphics[width=\textwidth]{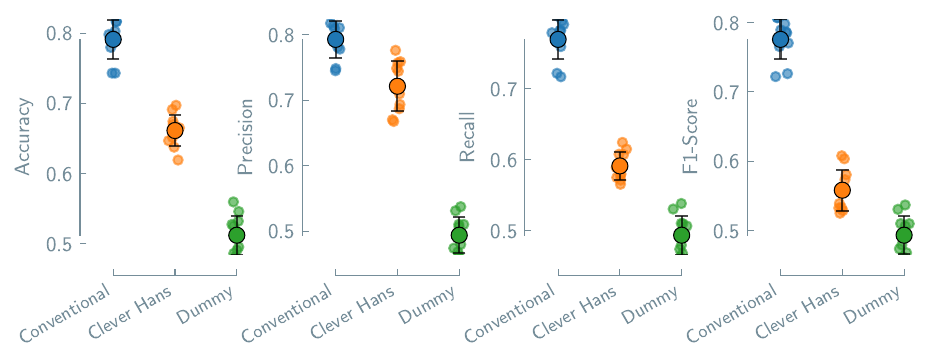}
	\label{fig:mof_solvent_removal_stability_metric_impact}
	\caption{\textbf{Performance of MOF solvent removal stability classification measured with different metrics.}  In all metrics, \enquote{Clever Hans models} outperform simple baselines. In some metrics, such as precision, \enquote{Clever Hans models} come close in performance to models directly trained on MOF descriptors (\enquote{Conventional}).}
	\script{analyze-mof-solvent-stability.py}
\end{figure}

\begin{figure}[htb]
	\includegraphics[width=\textwidth]{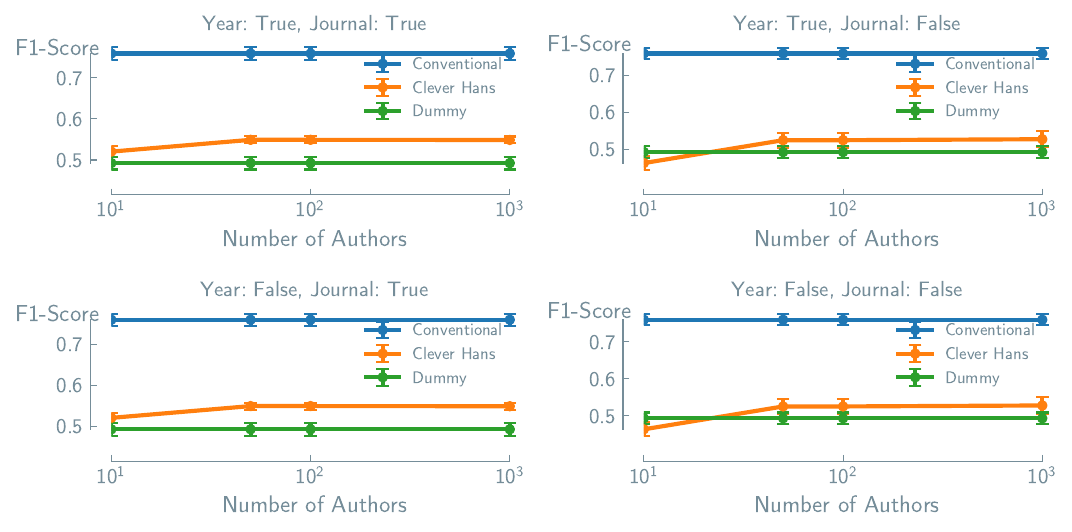}
	\caption{\textbf{Parameter sweep for MOF solvent stability.} Performance of proxy models as a function of the type and number of predicted bibliometric features.}
	\label{fig:mof_solvent_parameter_sweep}
	\script{analyze-mof-solvent-stability.py}	
\end{figure}

\Cref{fig:mof_solvent_removal_stability_metric_impact} shows the performance for MOF solvent removal stability classification measured with different metrics. 
\Cref{fig:mof_solvent_parameter_sweep} shows the performance of proxy models as a function of the type of bibliometric information used as model input.

\clearpage

\subsection{MOF Thermal Stability} \label{sec:appx-mof-thermal-stability}

\begin{figure}[htb]
	\includegraphics[width=\textwidth]{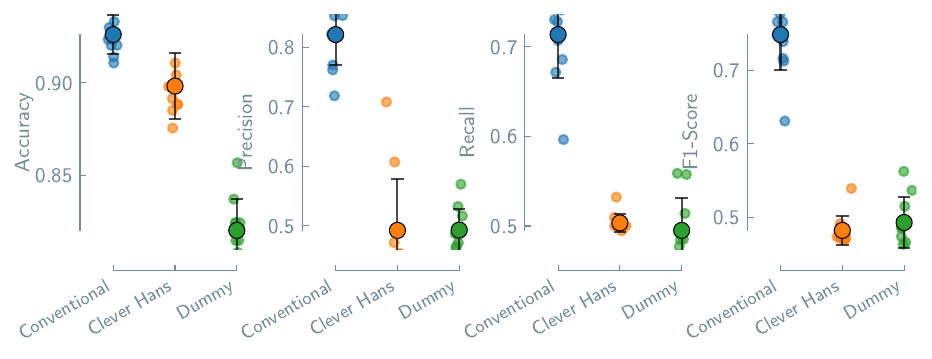}
	\caption{\textbf{MOF thermal stability prediction performance across different metrics.} The measured difference between conventional and proxy models depends on the evaluation metric chosen.}
	\label{fig:mof_thermal_stability_metric_impact}	
	\script{analyze-mof-thermal-top10.py}	
\end{figure}

\begin{figure}[htb]
	\includegraphics[width=\textwidth]{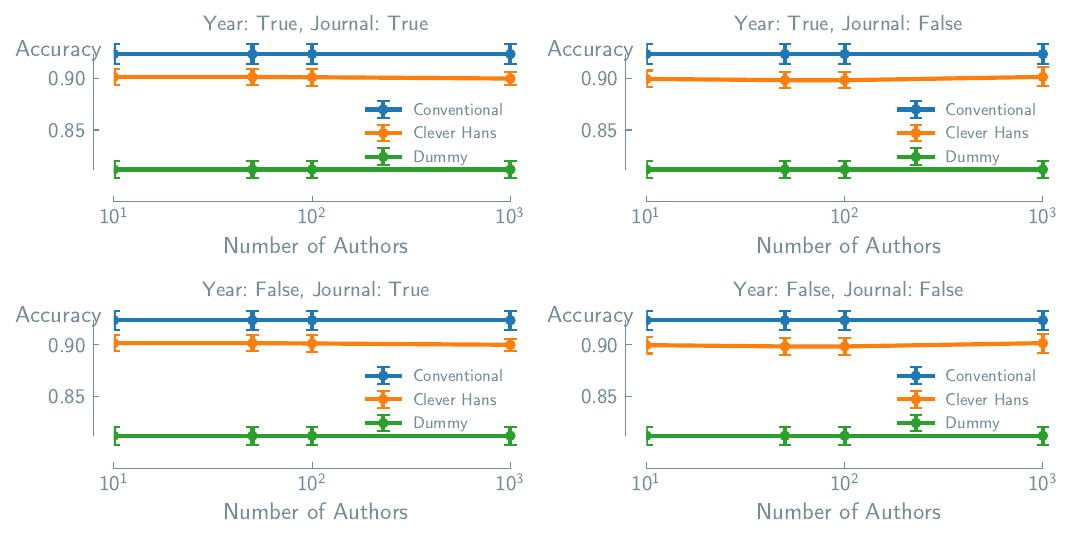}
	\caption{\textbf{Parameter sweep analysis for MOF thermal stability.} Performance of proxy models as a function of author count thresholds and inclusion of temporal/journal information.}
	\label{fig:mof_thermal_parameter_sweep}
	\script{analyze-mof-thermal-top10.py}
\end{figure}

\Cref{fig:mof_thermal_stability_metric_impact} shows that the measured difference in performance between models depends on the chosen metric.
\Cref{fig:mof_thermal_parameter_sweep} demonstrates how Clever Hans performance varies with the type and number of bibliometric features included.

\clearpage

\subsection{Perovskite Solar Cells: Regression Analysis}
\begin{figure}[htb]
	\centering
	\includegraphics[width=\textwidth]{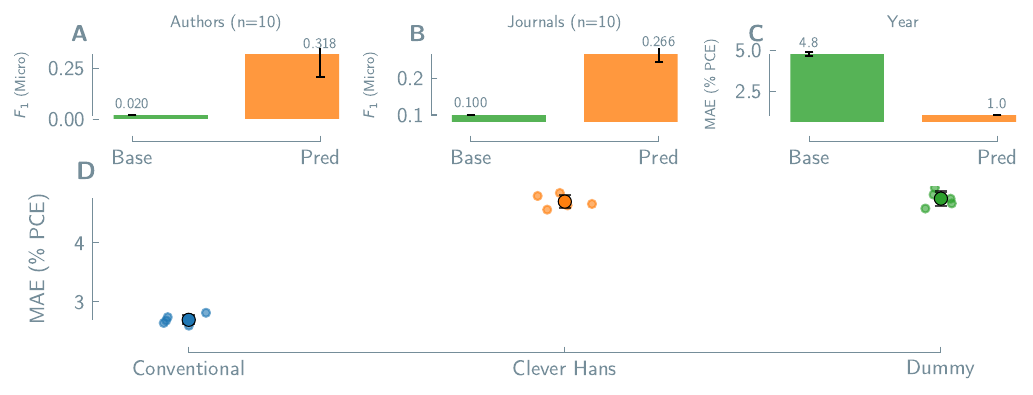}
	\caption{\textbf{Performance in predicting bibliometric information and in predicting photoconversion efficiencies.}}
	\label{fig:perovskite_main_regression}
	\script{analyze-perovskites.py}
\end{figure}

\begin{figure}[htb]
	\centering
	\includegraphics[width=\textwidth]{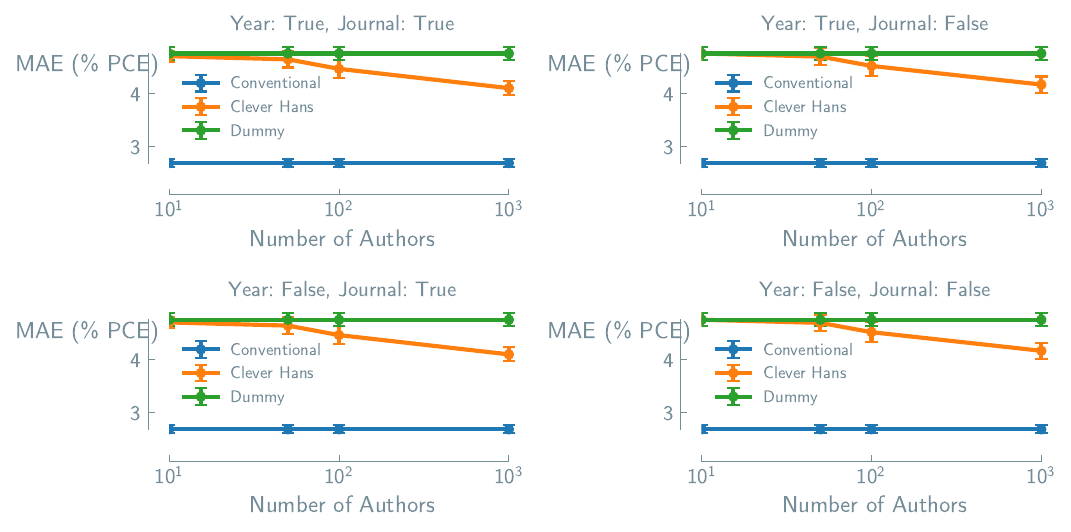}
	\caption{\textbf{Impact of the number of bibliometric features on the performance of \enquote{Clever Hans} models in predicting PCE.}}
	\label{fig:perovskite_parameter_sweep_regression}
	\script{analyze-perovskites.py}
\end{figure}

\begin{figure}[htb]
	\centering
	\includegraphics[width=\textwidth]{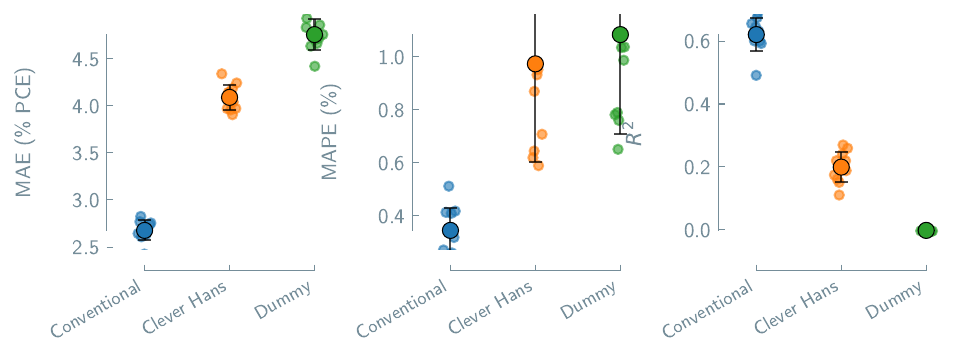}
	\caption{\textbf{Perovskite PCE prediction performance across metrics.} Comparison of conventional and proxy models using different regression evaluation metrics.}
	\label{fig:perovskite_performance_comparison_regression}
	\script{analyze-perovskites.py}
\end{figure}

\Cref{fig:perovskite_main_regression} shows the Clever Hans behavior for continuous PCE prediction (regression task). 
\Cref{fig:perovskite_parameter_sweep_regression} shows how performance varies with the number and type of bibliometric features.
\Cref{fig:perovskite_performance_comparison_regression} compares performance across different regression metrics. 

\clearpage

\subsection{Perovskite Solar Cells: Classification Analysis}
\begin{figure}[htb]
	\centering
	\includegraphics[width=\textwidth]{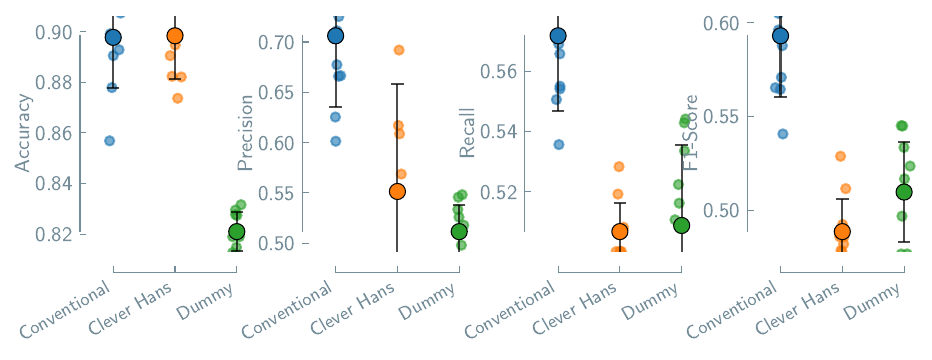}
	\caption{\textbf{Perovskite top-10\% classification performance across metrics.} Different evaluation metrics reveal varying degrees of Clever Hans effects.}
	\label{fig:perovskite_top10_performance_comparison}
	\script{analyze-perovskites-top10.py}
\end{figure}

\begin{figure}[htb]
	\centering
	\includegraphics[width=\textwidth]{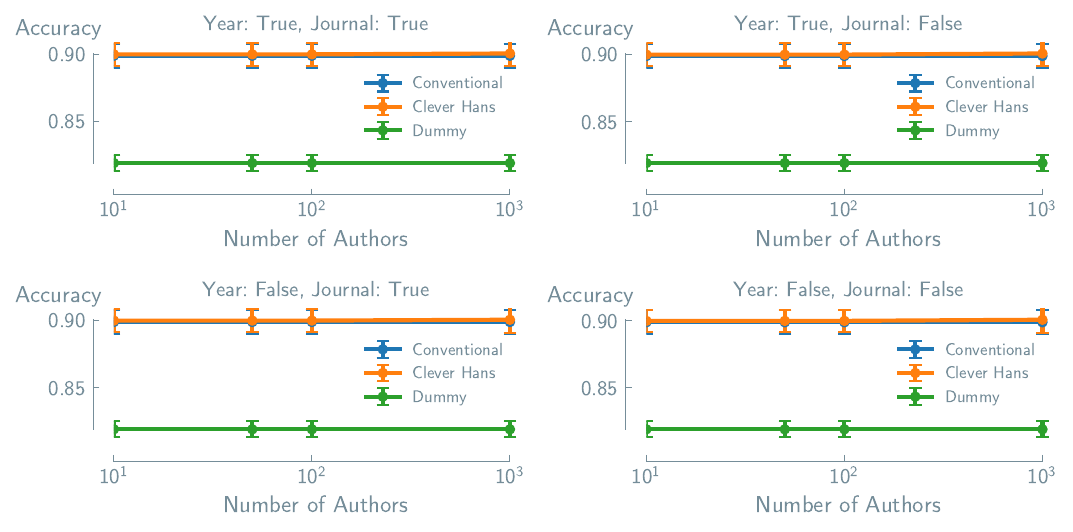}
	\caption{\textbf{Parameter sweep for perovskite top-10\% classification.} Performance sensitivity to author count thresholds and temporal/journal features.}
	\label{fig:perovskite_top10_parameter_sweep}
	\script{analyze-perovskites-top10.py}
\end{figure}

The classification analysis focuses on identifying top-performing devices rather than predicting exact efficiency values.
\Cref{fig:perovskite_top10_performance_comparison} shows metric-dependent effects in the classification setting.
\Cref{fig:perovskite_top10_parameter_sweep} demonstrates how classification performance varies with bibliometric feature selection.

\clearpage




\clearpage

\subsection{TADF Emitters}
\begin{figure}[htb]
	\centering
	\includegraphics[width=\textwidth]{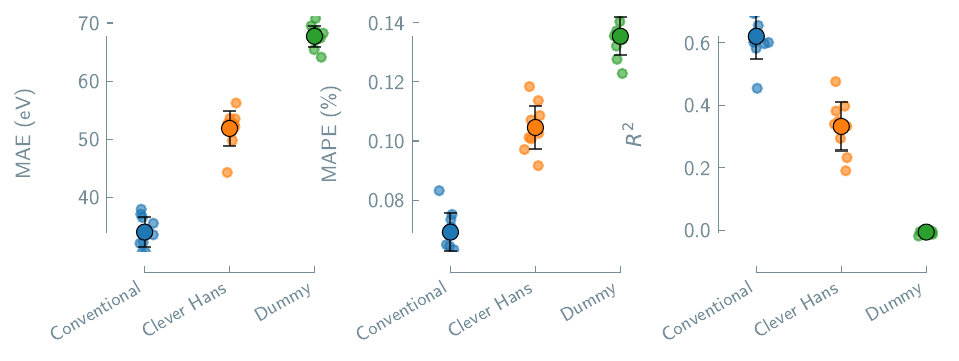}
	\caption{\textbf{TADF wavelength prediction performance across metrics.} Moderate proxy learning effects between conventional and baseline performance.}
	\label{fig:tadf_performance_comparison}
	\script{analyze-tadf.py}
\end{figure}

\begin{figure}[htb]
	\centering
	\includegraphics[width=\textwidth]{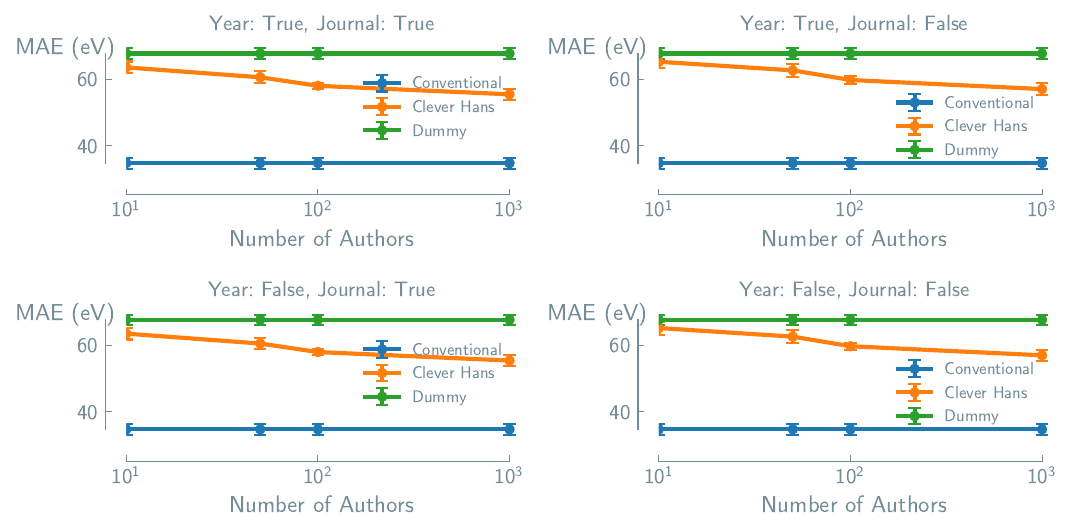}
	\caption{\textbf{Parameter sweep for TADF wavelength prediction.} Performance variation with different bibliometric feature configurations.}
	\label{fig:tadf_parameter_sweep}
	\script{analyze-tadf.py}
\end{figure}

TADF wavelength prediction shows intermediate Clever Hans effects, with proxy models performing between conventional and baseline approaches.
\Cref{fig:tadf_performance_comparison} compares performance across regression metrics.
\Cref{fig:tadf_parameter_sweep} demonstrates sensitivity to bibliometric feature selection.

\end{document}